# Cluster excitation and ionization in high velocity collisions: the atomic approach

F.Mezdari[1], K.Wohrer-Béroff [1*], M.Chabot[2]

[1]Laboratoire des Collisions Atomiques et Moléculaires CNRS UMR8625
Bâtiment 351, Université Paris-Sud 91405 Orsay Cedex, France
[2]Institut de Physique Nucléaire, IN2P3, CNRS, 91406 Orsay, Cedex, France
*corresponding author, Tel :+33 169157682 ; FAX : +33 169157671, E-mail : wohrer@lcam.u-psud.fr

**Abstract :**

The independent atom and electron model [1] is introduced in a quantum context and associated approximations tentatively estimated. Confrontation of the model to measured ionization and excitation cross sections of small ionic carbon clusters $C_n^+$ in collisions with helium at an impact velocity of 2.6 a.u is presented.

## I-Introduction

The treatment of excitation and ionization of clusters in high velocity collisions is a difficult task. A brief survey of theoretical approaches developed in the past has been published some years ago [2]. Since then, most of the theoretical studies have concerned stopping power calculations, which include the sum of excitation and ionization (see [3] and references therein). In the present model developed by the authors, a separate treatment of the two electronic processes is made. The model, referred as to the IAE model in the following, is based on a strong independent-atoms approximation [1]. In this paper, we try to estimate quantitatively the consequences of this "atomic approach" for predictions of integrated excitation and ionization cross sections (over final states, collision impact parameter and cluster orientation). For doing so, the IAE model is presented in a quantum context (&II) and associated approximations estimated for the $H_2$ molecule. In &III, confrontation between predictions of the model and experimental results for excitation and single and multiple ionization of small ionic carbon clusters is presented.

## II-The independent atom and electron collisional model

In this paragraph, we will formalize, give expressions and tentatively estimate the errors associated to the various approximations of the independent atom and electron collisional model. For simplicity, we consider the case of a collision between a structure less ion projectile and a cluster and assume a linear ion trajectory.

### 1-Quantum formalism

We start from the time dependent Schrödinger equation for the wave function of the cluster $\Psi(\vec{r}_i(\vec{R}_j),t)$ ($\vec{r}_i$ are positions of the electrons, depending on the positions $\vec{R}_j$ of nuclei). This equation is derived from the time independent Schrödinger equation in case the coupling between the system under study (here the cluster) and the environment (the impinging ion) is small [4]. We have (atomic units are used):

$$i\frac{d\Psi(\vec{r}_i,t)}{dt} = [H_S + H_I]\Psi(\vec{r}_i,t) \qquad (1)$$

$H_S$ is the hamiltonian of the cluster
$H_I$ is the interaction potential between the impinging ion (charge $Z_P$) and the cluster, depending on the impact parameter $\vec{b}$.

We decide to describe the state of the system (cluster) in the space E which is the tensorial product of the spaces of the N atomic constituents, which is always possible [5]. We have :
$$E = E_1 \otimes E_2 \otimes E_3 ..... \otimes E_N$$
and $|\Psi(t)\rangle$ writes, in an exact form :
$$|\Psi(t)\rangle = \sum_{j_1,j_2...j_N} a_{j_1 j_2...j_N}(t)|j_1(1)\rangle \otimes |j_2(2)\rangle \otimes ....... \otimes |j_N(N)\rangle \qquad (2)$$
where $|j_j(k)\rangle$ means atom k being in the state $j_j$.

The independent atom approximation contained in the IAE model consists in writing the initial wave function $\Psi(t=-\infty)$ as a tensorial product of atomic wave functions (usually ground states). We write :
$$|\Psi(t=-\infty)\rangle = |i_1(1)\rangle \otimes |i_2(2)\rangle \otimes ... \otimes |i_N(N)\rangle \equiv |i_1,i_2,...i_N\rangle \qquad (3)$$

On the other hand, $H_I$ writes :
$$H_I = \sum_{k=1->N} V_k(\vec{r}_k, \vec{b}_k, t) \qquad (4)$$
where $V_k$ is the projectile -atom k interaction potential and $b_k$ the impact parameter of incident ion with respect to atom k.

*α -first-order treatment*

Using (2)-(4) and solving (1) at a first order of perturbation theory, we find for the cluster wave function at finite time t :
$$\left|\Psi^{(1)}(t)\right\rangle \equiv [1 + c_{1i_1}(t) + c_{2i_2}(t) + ... + c_{Ni_N}(t)]|i_1,i_2,..i_N\rangle + \sum_{i\neq i_1} c_{1i}(t)e^{-iE_i t}|i,i_2,...i_N\rangle$$
$$+ \sum_{i\neq i_2} c_{2i}(t)e^{-iE_i t}|i_1,i,i_3,...i_N\rangle + .... + \sum_{i\neq i_N} c_{Ni}(t)e^{-iE_i t}|i_1,i_2,...i\rangle \qquad (5a)$$
where
$$c_{ni}(t) = \frac{1}{i}\int_{-\infty}^{t} e^{iE_i t'}\langle i|V_n(t')|i_n\rangle dt' \qquad (5b)$$
$$a_{ni}(t) = c_{ni}(t)e^{-iE_i t} \qquad (5c)$$

$a_{ni}(t)$ is the probability amplitude for the nth atom, initially in the state $|i_n>$, to be in the state $|i>$ of energy $E_i$ (with respect to ground state $|i_n>$) at time t. The cluster state is, in a first order perturbative treatment, a mixture of states having *a single atom excited*.

The probability $P_f(t)$ of finding the cluster in state $\Psi_f$ at time t is:

$$P_f(t) = |\langle \Psi_f | \Psi(t) \rangle|^2 \qquad (6)$$

as in the case of the initial state, we consider approximate final wave functions for the cluster which are tensorial products of atomic wave functions. Then it is readily found, using (5) and (6), that only states of the form $|E_1, i_2, \ldots i_N\rangle$, $|i_1, E_2, i_3 \ldots i_N\rangle, \ldots |i_1, i_2, \ldots E_N\rangle$ may be reached at a first order of the ion-cluster interaction potential. Since all these states are orthogonal, we have for the first order probability of finding the cluster in a state of excitation $E_f$ after the collision:

$$P_f^{(1)}(E_f, t=+\infty) = P_{at1}^{(1)}(E_f) + P_{at2}^{(1)}(E_f) + \ldots P_{atN}^{(1)}(E_f) \qquad (7)$$

where:

$$P_{atj}^{(1)}(E_j) = |c_{jE_j}(+\infty)|^2 = \left| \frac{1}{i} \int_{-\infty}^{+\infty} e^{iE_j t} \langle E_j | V_j(t) | i_j \rangle dt \right|^2 \qquad (8)$$

is the usual first –order atomic probability for promoting atom j in a state of excitation energy $E_j$ [5], a probability which depends on the impact parameter $b_j$ of incident ion with respect to atom j.

We deduce from what precedes that multiple electronic excitation of a cluster made of monoelectronic atoms cannot occur through a first order treatment. For double (triple…) excitation in this last case, a treatment to second (third…) order with respect to the perturbative potential must be done.

*β-second-order treatment*

The probability for double atom excitation of the cluster is derived in a second order treatment. For excitation of atoms i and j into states $E_i$ and $E_j$ respectively we find:

$$P_{ati,atj}^{(2)}(E_i, E_j) =$$

$$\left| \int_{-\infty}^{+\infty} e^{iE_j t} \langle E_j | V_j(t) | i_j \rangle \left\{ \int_{-\infty}^{t} e^{iE_i t'} \langle E_i | V_i(t') | i_i \rangle dt' \right\} dt + \int_{-\infty}^{+\infty} e^{iE_i t} \langle E_i | V_i(t) | i_i \rangle \left\{ \int_{-\infty}^{t} e^{iE_j t'} \langle E_j | V_j(t') | i_j \rangle dt' \right\} dt \right|^2 \qquad (9)$$

where the two terms in RHS of equation (9) express the possibility of exciting first atom i then atom j, or the opposite. Equation (9) may be formally written:

$$P_{ati,atj}^{(2)}(E_i, E_j) = \left| \int_{-\infty}^{+\infty} f(t) \left\{ \int_{-\infty}^{t} g(t') dt' \right\} dt + \int_{-\infty}^{+\infty} g(t) \left\{ \int_{-\infty}^{t} f(t') dt' \right\} dt \right|^2 \qquad (10a)$$

and it is found that, in case f=g, we have:

$$P_{ati,atj}^{(2)}(E_i, E_i) = \left| \left\{ \int_{-\infty}^{+\infty} f(t) \, dt \right\}^2 \right|^2 = (P_{ati}^{(1)}(E_i))^2 \qquad (10b)$$

Equation (10b) is then valid in case of two identical atoms, seen at the same impact parameter by the projectile and excited in the same final state. In case two identical atoms are seen at different impact parameters and are excited in the same final state, there is a compensation between the two terms or RHS of equation (10a) when considering various azimuthal angles of $\vec{b}$ and we can write with a good approximation:

$$P_{ati,atj}^{(2)}(E_i, E_j) = P_{ati}^{(1)}(E_i) \, P_{atj}^{(1)}(E_j) \qquad (11)$$

In the case of double electronic excitation (ionization) of a cluster, two routes are then generally possible : either by excitation (ejection) of two electrons from the same atom using equation (7) with the probability (8) being the product of individual electron probabilities within the independent electron approximation (we do not comment further this well documented approximation [6 and references therein]); or, by excitation (ionization) of one electron on two different atoms, using equation (11). Within second order treatment, we find also that the probability of exciting atom i without exciting atom j has the form of equation (11) with $P_{atj}^{(1)}(E_j)$ replaced by $(1- P_{atj}^{(1)}(E_j))$. We retrieve then, for single and double ionization, equations reported in previous references within the IAE model [1,7] ; the case of triple ionization is derived in a similar manner and includes first-order, second-order and triple-order contributions.

2-Validity of the approximations

Whereas the independent electron approximation has been applied with success, even in outer shells of atoms [6], the most severe approximation of the IAE model is the assumption of independent atoms within the cluster. In particular, writing (3) in place of (2) for the initial cluster wave function has the following qualitative consequences:

i) the electronic density, electronic binding energies are incorrect
ii) neglect of the sharing of electrons between atoms, a typical « molecular » effect, will destroy interference effects on excitation probabilities

We will discuss these points with reference to the most simple case, *single ionisation of an hydrogen molecule* by impact of a bare ion, presently the subject of great experimental and theoretical interests ([8] and references therein).

Within the IAE model, $H_2$ is made of two independent H atoms, whose states coincide with the state of the unique electron (electron 1 for atom 1, electron 2 for atom2). We express the probability of ionisation of $H_2$ between the initial state $\Psi_{i-IAE}$ (ground state) and the final state $\Psi_{f-IAE}$ (plane wave $\vec{k}$ for one of the electron). We have, according to what precedes:

$|\Psi_{i-IAE}\rangle = |1s(1), 1s(2)\rangle$ (12a)
$|\Psi_{f-IAE}\rangle = |k(1), 1s(2)\rangle$ and $|1s(1), k(2)\rangle$ (12b)
$P_{SI-IAE}^{(1)}(\vec{k},\vec{b}) = |c_{1k}(+\infty, b_1)|^2 + |c_{2k}(+\infty, b_2)|^2$ (13)

where $b_1$ and $b_2$ depend on $\vec{b}$ and on the molecular orientation [1].

On the other hand, the simplest electronic molecular wave function for the ground state of $H_2$ is of the Heitler-London form (omitting the spin wavefunction) [9,10]:

$$\left|\Psi_{i-moleculeH_2}\right\rangle = \frac{1}{\sqrt{2(1+S^2)}}[\left|1s^\lambda(1), 1s^\lambda(2)\right\rangle + \left|1s^\lambda(2), 1s^\lambda(1)\right\rangle]$$ (14a)

where $1s^\lambda(i)$ are 1s hydrogenic wavefunctions centered on nuclei i with charge $\lambda$ and S is the overlap integral of the functions on the two centres.

The orbital symmetric final state for single ionisation of $H_2$ (k relative to the centre of the molecule) writes:

$$|\Psi_{f-moleculeH_2}\rangle = \frac{1}{2\sqrt{(1+S)}}[|k,1s^\lambda(1)\rangle + |k,1s^\lambda(2)\rangle + |1s^\lambda(1),k\rangle + |1s^\lambda(2),k\rangle] \quad (14b)$$

Using equations (14) we find for the first-order probability of single ionisation of $H_2$ (formula (8) applied to molecular states):

$$P_{SI-moleculeH_2}^{(1)}(\vec{k},\vec{b}) = \frac{(1+S)}{2(1+S^2)}\left|\int_{-\infty}^{+\infty} e^{iE_{fi}t}dt[e^{i\vec{k}\cdot\vec{\rho}/2}\langle k(1)|V(t)|1s^\lambda(1)\rangle + e^{-i\vec{k}\cdot\vec{\rho}/2}\langle k(2)|V(t)|1s^\lambda(2)\rangle]\right|^2$$

(15)

where $\vec{\rho}$ is the internuclear distance between atoms 1 and 2 ($\rho$=1.4 au in $H_2$).

In (15) $E_{fi}$ is the energy difference between the initial and final molecular states. If we assume this energy difference to be equal to the energy difference within the atoms, then we may express (15) as a function of the atomic probability amplitudes $c_{jk}$ defined previously (formula (5b):

$$c_{1k}^\lambda(+\infty,b_j) = \frac{1}{i}\int_{-\infty}^{+\infty} e^{iE_k t}\langle k(j)|V(t)|1s^\lambda(j)\rangle dt \quad (16)$$

and write the molecular single ionisation probability as:

$$P_{SI-moleculeH_2}^{(1)}(\vec{k},\vec{b}) = \frac{(1+S)}{2(1+S^2)}\left|e^{i\vec{k}\cdot\vec{\rho}/2}c_{1k}^\lambda(+\infty,b_1) + e^{-i\vec{k}\cdot\vec{\rho}/2}c_{2k}^\lambda(+\infty,b_2)\right|^2$$

$$= \frac{(1+S)}{2(1+S^2)}\left\{|c_{1k}^\lambda(+\infty,b_1)|^2 + |c_{2k}^\lambda(+\infty,b2)|^2 + 2\text{Re}(e^{i\vec{k}\cdot\vec{\rho}}c_{1k}^\lambda(+\infty,b_1)c_{2k}^\lambda(+\infty,b_2)^*)\right\}$$

(17)

Crossed terms in equation (17) depend clearly on $\vec{k}$, $\vec{u}$ (the molecule direction) and $\vec{b}$. A simple case arises when the molecule is aligned along the beam ($\theta$=0°); then $\vec{b1}=\vec{b2}=\vec{b}$ and we obtain:

$$P_{SI-moleculeH_2}^{\Theta=0°(1)}(\vec{k},\vec{b}) = \frac{(1+S)}{2(1+S^2)}4|c_{1k}^\lambda(+\infty,b)|^2 \cos^2(\vec{k}\cdot\vec{\rho}/2) \quad (18)$$

for $\vec{k}\perp\vec{\rho}$ or for $k_{//} \leq 0.5$ a.u ($E_{e^-} \leq 3$eV), we have $\cos^2(\vec{k}\cdot\vec{\rho}/2) \sim 1$ i.e interferences are constructive. Note that these conditions are those reached predominantly in high velocity collisions [11]. Using $\lambda$=1.17 [10] and $S(\lambda)$=0.67 [12], we find:

$$P_{SI-moleculeH_2}^{\Theta=0°(1)}(\vec{k},\vec{b}) \approx 2.3|c_{1k}^\lambda(+\infty,b)|^2 = 1.15 P_{SI-IAE}^{(1)(\lambda)}(\vec{k},b) \quad (19)$$

that means, and since the screening parameter tends to reduce the probability, a value close to what is obtained within the independent atom approximation (formula (13)).

When $\vec{u}$ makes an angle $\theta\neq 0°$ with the incident beam, interference patterns are smaller [8]. In the end, integrating (19) over $\vec{b}$ we find for the ionisation cross section in $H_2$ roughly twice the ionization cross section in H, as observed experimentally at high enough energy [13].

The case of more complex systems will not be treated at all and will be the subject of future work. One can expect that, for the small carbon clusters which have been experimentally studied and whose comparison with the IAE model is given in the next paragraph, these approximations remain reasonable. In carbon clusters, the electronic wave function of the ground state has often been developed on a linear combination of atomic 2s and 2p orbitals only, either in the first studies of these cluster structure and relative stability [14] or in Tight Binding Molecular Dynamics simulations [15]. Also, the standard of precision is reduced when interested in integrated quantities (over $\vec{b}$, over the final states of ionization or excitation, over the orientation of the object) as seen before in the case of $H_2$.

**III-Confrontation to experiment**

We compare predictions of the model to experiments performed with small ionic carbon clusters $C_n^+$ (n≤5) colliding with helium atoms at a constant impact velocity of 2.6 atomic units (au). The experiments have been performed at the Tandem facility in Orsay (France) with beams of accelerated clusters. All details about the experimental set-up and analysis method have been given in previous papers [1,16].

1-Ionization cross sections of small ionic carbon clusters

Experimental cross sections are presented in figure 1 as a function of the cluster size n, together with predictions of the IAE model. As explained in [17], various lines corresponding to different normalisations of the theory to the experimental $C^+$->He point are presented. The overall size dependence of the cross sections are relatively well reproduced by the model in all cases, in particular the large enhancement of the double ionization cross section when going from n=1 to n=2.

2-Excitation cross sections of small ionic carbon clusters; energy deposit

In figure 2 results concerning electronic excitation of $C_n^+$ in the same systems are presented: experimental cross sections refer to the dissociative part of the cross section whereas the IAE model is done for the total excitation cross section. As for ionization, classical trajectory Monte Carlo (CTMC) calculations were performed for excitation of the $C^+$ ion. Contrary to ionization, no measurement exists yet for the excitation of $C^+$ in the $C^+$->He collision so that this theoretical calculation was introduced directly into the model in order to predict excitation of heavier systems (solid line). Here again, a reasonable agreement is found. In order to interpret the fragmentation of the excited clusters ([1,18] and to be published), a calculation of energy deposit has been performed within the IAE model, which is presented in figure 3 for the case of $C_5^+$. Note that we calculate here, contrary to what is derived in stopping power calculations, the energy due to electronic excitation only (without ionization). Interpretation of the experimental fragmentation data will associate this energy distribution (to which 3eV of vibrational energy of the cluster in the entrance channel has to be added [16]) to predictions of a statistical fragmentation theory [19].

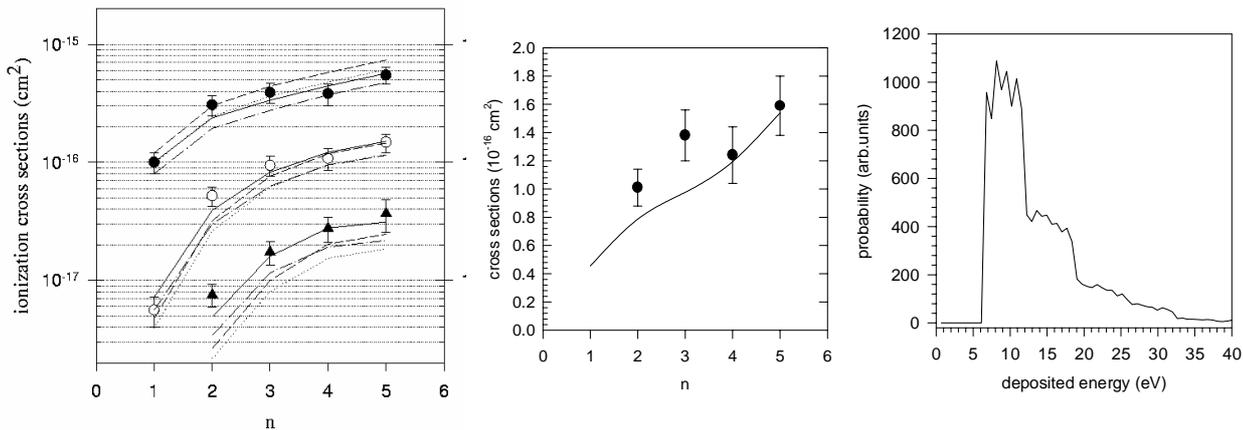

*Figure1 (left): single (black circles), double (open circles) and triple (black triangles) ionisation cross sections of $C_n^+$ clusters as a function of n, and results of the IAE model (see text)*
*Figure 2 (medium) : experimental and calculated excitation cross sections of $C_n^+$ clusters as a function of n*
*Figure 3 (right): calculated energy deposit due to electronic excitation in $C_5^+ \rightarrow He$*